\documentclass[twocolumn,showpacs,superscriptaddress,amsmath,amssymb]{revtex4}
\usepackage{graphicx}% Include figure files
\usepackage{dcolumn}% Align table columns on decimal point
\usepackage{bm}% bold math
\usepackage{color}
\def\bd{
\begin{document}} \def\ed{\end{document}}
\def\bmp{\begin{minipage}} \def\emp{\end{minipage}}
\def\bcc{\begin{center}} \def\ecc{\end{center}}     \def\npg{\newpage}
\def\beq{\begin{equation}} \def\eeq{\end{equation}} \def\hph{\hphantom}
\def\be{\begin{equation}} \def\ee{\end{equation}} \def\r#1{$^{[#1]}$}
\def\n{\noindent} \def\ni{\noindent} \def\pa{\parindent}
\def\hs{\hskip} \def\vs{\vskip} \def\hf{\hfill} \def\ej{\vfill\eject}
\def\cl{\centerline} \def\ob{\obeylines}  \def\ls{\leftskip}
\def\underbar#1{$\setbox0=\hbox{#1} \dp0=1.5pt \mathsurround=0pt
   \underline{\box0}$}   \def\ub{\underbar}    \def\ul{\underline}
\def\f{\left} \def\g{\right} \def\e{{\rm e}} \def\o{\over} \def\d{{\rm d}}
\def\vf{\varphi} \def\pl{\partial} \def\cov{{\rm cov}} \def\ch{{\rm ch}}
\def\la{\langle} \def\ra{\rangle} \def\EE{e$^+$e$^-$} \def\pt{p_{\rm t}}
\def\pti{p_{{\rm t},i}} \def\vti{v_{{\rm t},i}}
\def\ptj{p_{{\rm t},j}}\def\Pt{P_{\rm t}} \def\vt{v_{\rm t}}

\def\bitz{\begin{itemize}} \def\eitz{\end{itemize}}
\def\btbl{\begin{tabular}} \def\etbl{\end{tabular}}
\def\btbb{\begin{tabbing}} \def\etbb{\end{tabbing}}
\def\beqar{\begin{eqnarray}} \def\eeqar{\end{eqnarray}}
\def\\{\hfill\break} \def\dit{\item{-}} \def\i{\item}
\def\bbb{\begin{thebibliography}{9} \itemsep=-1mm}
\def\ebb{\end{thebibliography}} \def\bb{\bibitem}
\def\bpic{\begin{picture}(260,240)} \def\epic{\end{picture}}
\def\akgt{\cl{\bf ACKNOWLEDGMENTS}}
\def\fgn{\noindent{\bf\large\bf figure captions}}
%%%%%%%%%%%%%%%%%%%%%%%%%%%%%%%%%%%%%%%%%%%%%%%%%%%%%%%%%%%%%%%%%%%%%%
\def\m1{\langle N_p\rangle} \def\u2{\langle N_{\bar p}\rangle} \def\Nap{N_{\bar
p}}
%%%%%%%%%%%%%%%%%%%%%%%%%%%%%%%%%%%%%%%%%%%%%%%%%%%%%%%%%%%%%%%%%%%%%%%%%%%%%%
\def\lan{\langle}
\def\ran{\rangle}
\def\p{\pi}
\def\ifmath#1{\relax\ifmmode #1\else $#1$\fi}%
\def\rc{\ifmath{{\mathrm{c}}}}
\def\cut{\ifmath{{\mathrm{cut}}}}
\def\rF{\ifmath{{\mathrm{F}}}}
\def\rK{\ifmath{{\mathrm{K}}}}
\def\rp{\ifmath{{\mathrm{p}}}}
\def\rt{\ifmath{{\mathrm{t}}}}
\def\LAB{\ifmath{{\mathrm{LAB}}}}
\def\cut{\ifmath{{\mathrm{cut}}}}
\def\beq{\begin{equation}}
\def\eeq{\end{equation}}
\newcommand{ \eps }{{\varepsilon}}
\newcommand{ \rar }{\rightarrow}
\newcommand{ \lar }{\leftarrow}
\newcommand{ \as }{$\alpha_{s}$}
\newcommand{ \pT }{\mbox{$p_{T}$}}
\newcommand{ \dzero}{$D^{0}$}
\newcommand{ \vv }{$v_{2}$}
\newcommand{\sqrts}{\mbox{$\sqrt{s}$}}
\newcommand{\sNN}{{{$\sqrt{s_{_{{NN}}}}$}}}
\newcommand{\dNdeta}{\mbox{$dN_{\mathrm{ch}}/d\eta$}}
\newcommand{\dAu}{{\textit{d}+Au}}
\newcommand{\AuAu}{Au+Au}
\newcommand{\pp}{\mbox{$p+p$}}
\newcommand{\mev}{\mbox{$\mathrm{MeV}$}}
\newcommand{\gev}{\mbox{$\mathrm{GeV}$}}
\newcommand{\gevcc}{\mbox{$\mathrm{GeV/}c^2$}}
\newcommand{\mevcc}{\mbox{$\mathrm{MeV/}c^2$}}
\newcommand{\gevc}{\mbox{${\mathrm{GeV/}}c$}}
\newcommand{\mevc}{\mbox{${\mathrm{MeV/}}c$}}
\newcommand{\TAA}{\mbox{$\mathrm{T}_{AA}$}}
\newcommand{\raa}{\mbox{$R_{AA}$}}
\newcommand{\RAA}{\mbox{$R_{AuAu/dAu}$}}
\newcommand{\RCP}{\mbox{$R_{CP}$}}
\newcommand{\etc}{\mbox{\textit{etc.}}}
\newcommand{\dedx}{\mbox{$dE/dx$}}
\newcommand{\muBT}{\mbox{$\mu_{B}/T$}}
\newcommand{\muB}{\mbox{$\mu_{B}$}}
\newcommand{\npart}{\mbox{$N_{\mathrm{part}}$}}
\newcommand{\nbin}{\mbox{$N_{\mathrm{bin}}$}}
\newcommand{ \jpsi }{$J/\psi$}
\newcommand{\KV}{{\mbox{$\kappa\sigma^{2}$}}}
\newcommand{\SD}{{\mbox{$S\sigma$}}}
\newcommand{\KDS}{{\mbox{$\kappa \sigma/S$}}}
\newcommand{\VM}{{\mbox{$\sigma^{2}/M$}}}

\newcommand{\ka}{{\mbox{$\kappa$}}}
\newcommand{\si}{{\mbox{$\sigma$}}}
\newcommand{\var}{{\mbox{$\sigma^{2}$}}}
\newcommand{\tc}{{\mbox{$T_c$}}}

\usepackage[colorlinks,
           bookmarks=true,
           linkcolor=blue,
           %pdfborder=blue?
           urlcolor=blue,
            anchorcolor=black,
            citecolor=blue
            ]{hyperref}
            
\usepackage[all]{hypcap}

\newcommand{\cinst}[2]{$^{\mathrm{#1}}$~#2\par}
\newcommand{\crefi}[1]{$^{\mathrm{#1}}$}
\newcommand{\crefii}[2]{$^{\mathrm{#1,#2}}$}
\newcommand{\crefiii}[3]{$^{\mathrm{#1,#2,#3}}$}
\newcommand{\HRule}{\rule{0.5\linewidth}{0.5mm}}

\bd
\title{Cumulants of Net-Proton, Net-Kaon and Net-Charge Multiplicity Distributions in Au+Au Collisions at RHIC BES Energies from UrQMD Model}

\author{Ji Xu}
\affiliation{ Key Laboratory of Quark \& Lepton Physics (MOE) and Institute of Particle Physics, Central China Normal University, Wuhan 430079, China}
\affiliation{Lawrence Berkeley National Laboratory, Berkeley, CA 94720, USA}
\author{Shili Yu}
\author{Feng Liu}
\author{Xiaofeng Luo }
\email{xfluo@mail.ccnu.edu.cn}
\affiliation{ Key Laboratory of Quark \& Lepton Physics (MOE) and Institute of Particle Physics, Central China Normal University, Wuhan 430079, China}

\begin{abstract}
Fluctuations of conserved quantities are sensitive observables to probe the signature of QCD phase transition and critical point in heavy-ion collisions.  With the UrQMD model, we have studied the centrality and energy dependence of various order cumulants and cumulant ratios (up to fourth order) of net-proton,net-charge and net-kaon multiplicity distributions in Au+Au collisions at  $\sqrt{s_{NN}}$= 7.7, 11.5, 19.6, 27, 39, 62.4, 200 GeV. The model results show that the production mechanism of the particles and anti-particles have significant impacts on the cumulants of net-particles multiplicity distributions and show strong energy dependence. We also made comparisons between model calculations and experimental data measured in the first phase of the beam energy scan (BES) program by the STAR experiment at RHIC. The comparisons indicate that the baryon conservation effect strongly suppress the cumulants of net-proton distributions at low energies and the non-monotonic energy dependence for the net-proton {\KV} at the most central Au+Au collisions measured by the STAR experiment can not be described by the UrQMD model. Since there has no physics of QCD phase transition and QCD critical point implemented in the UrQMD, the model results provide us baselines and qualitative estimates about the non-critical background contributions to the fluctuations observables in heavy-ion collisions. 
\end{abstract}

%\pacs{25.75.Gz,12.38.Mh,21.65.Qr,25.75.-q,25.75.Nq}
%\begin{keyword}
%keyword,  3--8 words separated by comma
%\end{keyword}

%%%%%%%%%%%%%%%%%%% (1) Introducion  %%%%%%%%%%%%%%%%%%%%%%%%

\maketitle
\section{Introduction}
One of the main goals of the high energy nuclear collisions is to explore the phase structure of the strongly interacting hot and dense nuclear matter. The QCD phase structure can be displayed in the two dimensional (temperature ($T$) Vs. baryon chemical potential ($\mu_{B}$)) QCD phase diagram. The first principle Lattice QCD calculations demonstrate that the transition from the hadron phase to Quark-Gluon Plasma (QGP) at zero $\mu_{B}$ is a smooth crossover~\cite{ref1}. While at large $\mu_{B}$ region, the phase transition could be of the first order~\cite{ref2}. Thus, there should be a QCD Critical Point (CP) as the end point of the first order phase boundary towards the crossover region \cite{ref3}. Experimental discovery of the critical point will be a landmark for the study of the phase structure of QCD matters.

Fluctuations of conserved quantities, such as net-baryon (B), net-charge (Q) and net-strangeness (S),  have long been predicted to be sensitive to the QCD phase transition and QCD critical point. Experimentally, one can measure various order moments (Variance($\sigma^2$), Skewness($S$), Kurtosis($\kappa$)) of the event-by-event conserved quantities distributions in heavy-ion collisions. These moments are sensitive to the correlation length ($\xi$) of the hot dense matter created in the heavy-ion collisions~\cite{ref4} and also connected to the thermodynamic susceptibilities computed in Lattice QCD \cite{ref3,ref5,ref6} and in the Hadron Resonance Gas (HRG) \cite{ref7,ref7.1,ref7.2,ref7.3} model. 
For instance, the variance, skewness and kurtosis have been shown to be related to the different power of the correlation length as $\xi^2$, $\xi^{4.5}$ and $\xi^7$ \cite{ref4}, respectively. Theoretically,  the $n^{th}$ order susceptibilities $\chi ^{(n)}$ are related to cumulant as ${\chi ^{(n)}} = C_n^{}/V{T^3}$~\cite{ref9}, where $V,T$ are the volume and temperature of the system, $C_{n}$ is the $n^{th}$ order cumulants of multiplicity distributions. In order to compare with the theoretical calculations, cumulant ratios ($S\sigma=C_{3}/C_{2}, \kappa\sigma^2=C_{4}/C_{2}$) are constructed to cancel the volume effects. Thus, those moment products are also directly related to the ratios of various order susceptibilities as $\kappa\sigma^2$=$\chi_{B}^{(4)}/\chi_{B}^{(2)}$ and $S\sigma$=$\chi_{B}^{(3)}/\chi_{B}^{(2)}$. Due to the high sensitivity to the correlation length and the connection with the susceptibilities, one can use the moments of the conserved quantity distributions to search for the QCD critical point and the QCD phase transition. These have been widely studied experimentally and theoretically~\cite{ref4,ref7,ref9,ref10,ref11,ref12,2014_Bengt_flu,2015_JianDeng_fluctuation,2016_fluctuations_Huichao}. To locate the critical point and map out the first order phase boundary,  the first phase of the beam energy scan program has started in the year 2010 at RHIC. It tunes the Au+Au colliding energies from 200 GeV down to 7.7 GeV~\cite{ref12.1,BESII_WhitePaper}, which correspond to a baryon chemical potentials range from 20 to 420 MeV. In this paper, we have studied the centrality and energy dependence for the moments of the net-proton, net-charge, net-kaon multiplicity distributions in Au+Au collisions at $\sqrt{s_{NN}}$ = 7.7, 11.5, 19.6, 27, 39, 62.4 and 200GeV with the UrQMD model with version 2.3~\cite{ref14}. These results are used to understand how the various non-critical physics process, like hadronic re-scattering and conservation law, contribute to the moments observables in heavy-ion collisions. It provides us the baselines and qualitative estimates on the background effects for the experimental search for the QCD phase transition and QCD critical point. 

The paper is organized as follows. In section II, we briefly introduce the UrQMD model. In the section III, we provide the mathematical definition for the observables used in the data analysis. Then, we present cumulants ($C_{1},C_2,C_3,C_4$) of net-proton, net-charge and net-kaon multiplicity distributions in Au+Au collisions at RHIC BES energies from UrQMD model in the section IV. In section V, we made the comparisons between the model results and the measurements from the STAR experiments. Finally, summary is given in the section VI.

%%%%%%%%% %%%%%%%%%% (2) UrQMD Model %%%%%%%%%%%%%%%

\section{UrQMD Model}
The Ultra Relativistic Quantum Molecular Dynamics (UrQMD) is a microscopic many-body approach to study p + p, p + A, and A + A interactions at relativistic energies and is based on the covariant propagation of color strings, constituent quarks, and diquarks accompanied by mesonic and baryonic degrees of freedom. Furthermore it includes rescattering of particles, the excitation and fragmentation of color strings, and the formation and decay of hadronic resonances. UrQMD is a transport model for simulating heavy-ion collisions in the energy range from SIS (SchwerIonen Heavy-ion Synchrotron) to Relativistic Heavy Ion Collider (RHIC) and even in the Large Hadron Collider (LHC). It combines different reaction mechanism, and can provide theoretical simulated results of various experimental observables. The main parts of UrQMD model are: two body reaction cross section, two body potential and decay width. More details about the UrQMD model can be found in the reference~\cite{ref14,UrQMD}.

In this paper, we performed our calculations with UrQMD model in version 2.3 for Au+Au collision at $\sqrt{s_{NN}}$=7.7, 11.5, 19.6, 27, 39, 62.4, 200 GeV and the corresponding statistics are 72.5, 105, 106, 81, 133, 38, 56 million, respectively.

%%%%%%%%%%%%%%%%%%% (3) Observables %%%%%%%%%%%%%%%%%

\section{Observables}
In statistics~\cite{ref14.1}, the distribution function can be characterized by the various order moments, such as mean (M), variance ($\sigma^2$), skewness ($S$), kurtosis ($\kappa$). Before introducing the above moments used in our analysis, we would like to define cumulants, which are alternative approach compared to moments to characterize a distribution. 
%The cumulants determine the moments in the sense that any two probability distributions whose cumulants are identical will have identical moments as well and similarly the moments can determine the cumulants.

In the moments analysis, we use $N$ to represent the net-proton ($N_{p} - N_{\bar{p}}$), net-charge ($N_{+} - N_{-}$) and net-kaon($N_{K^{+}} - N_{K^{-}}$) number in one event. The average value over whole event ensemble is denoted by $<N>$. We use $\delta N= N- < N >$ to denote the deviation of $N$ from its mean value. Then the various order cumulants of event-by-event distributions of a variable N are defined as:
   
     \begin{eqnarray}
     &&C_{1,N} = < N > \\                                           
     &&C_{2,N} = < (\delta N)^2 >\\                             
     &&C_{3,N} = < (\delta N)^3 > \\                          
     &&C_{4,N} = < (\delta N)^4 >-3 < (\delta N)^2 >
     \end{eqnarray}
   
%An important properties of the cumulants is their additivity for independent variables. If $X$ and $Y$ are two independent random variables, then we have $C_{i,X+Y} = C_{i,X} + C_{i,Y}$ for $i$th order cumulant. We will use this property in our study.

Once we have the definition of cumulants, various moments can be denotes as:
\begin{eqnarray}
M = C_{1,N}, \sigma^2 = C_{2,N}, S=\frac{C_{3,N}}{(C_{2,N})^\frac{3}{2}}, \kappa = \frac{C_{4,N}}{(C_{2,N})^2}
\end{eqnarray}
In addition, the moments product $\kappa\sigma^2$ and $S\sigma$ can be expressed in term of cumulant ratio:
\begin{eqnarray}
\label{eq6} \kappa\sigma^2 = \frac{C_{4,N}}{C_{2,N}}, S\sigma = \frac{C_{3,N}}{C_{2,N}}
\end{eqnarray}
With above definitions, we can calculate various cumulants and ratios of cumulants for the measured event-by-event net-proton,net-charge and net-kaon multiplicity distributions. For the statistical error calculations, we use the formula that derived from the standard error propagation, which is based on the Delta theorem. Since we are focusing on the non-CP physics contribution to the fluctuation observable in heavy-ion collisions,  the efficiency 
effects are not studied in this paper with the UrQMD model simulations. The statistical errors of various order cumulants ($C_{n}$) can be approximated as $error(C_{n}) \propto \sigma^{n}/(\epsilon^{n}\sqrt{N})$, where $\sigma$ is the width of the measured distribution, $\epsilon$ is the detection efficiency of the measured particles and N is the number of events. More detail discussion about the efficiency correction and error estimation can be found in the references~\cite{ref18,ref18_1}.

%%%%%%%%%%%%%%%%%%%% (4) Model Results %%%%%%%%%%%%%%%%%%%%

\section{Results from UrQMD Model}
In the UrQMD calculations, we applied the same kinematic cuts as used in the data analysis \cite{ref17.1}.  The protons and anti-protons are obtained at mid-rapidity $(|y|<0.5)$ and within the transverse momentum $0.4<p_T<0.8$ GeV/c, the charged particles in the net-charge fluctuations are measured at pseudo-rapidity range $|\eta|<0.5$ and within the transverse momentum $0.2<p_T<2$ GeV/c, and the $K^{+}$ ( $K^{-}$ ) for net-kaon study are measured at mid-rapidity $(|y|<0.5)$ within the transverse momentum $0.2<p_T<1.6$ GeV/c. 
\begin{figure*}
\hspace{-2.0cm}
\includegraphics[width=6.in]{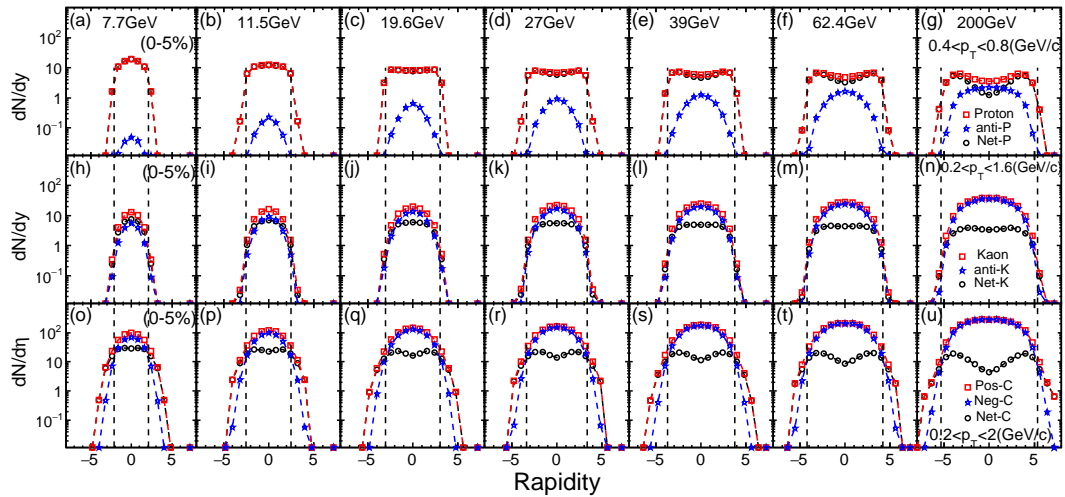}
\caption{\label{Fig_1} (Color online) The $dN/dy$ distributions of net-proton, proton, anti-proton, net-kaon, $K^{+}$, $K^{-}$ and $dN/d\eta$ distributions of net-charge, positive and negative charge multiplicity in Au+Au collisions at $\sqrt{s_{NN}}$ =7.7, 11.5,19.6, 27, 39, 62.4 and 200GeV for most central collision (0-5\%). The vertical black dashed lines in the figure represent the beam rapidity for each energy. }
\end{figure*}

\begin{figure*}
\hspace{-2.0cm}
\includegraphics[width=6.in]{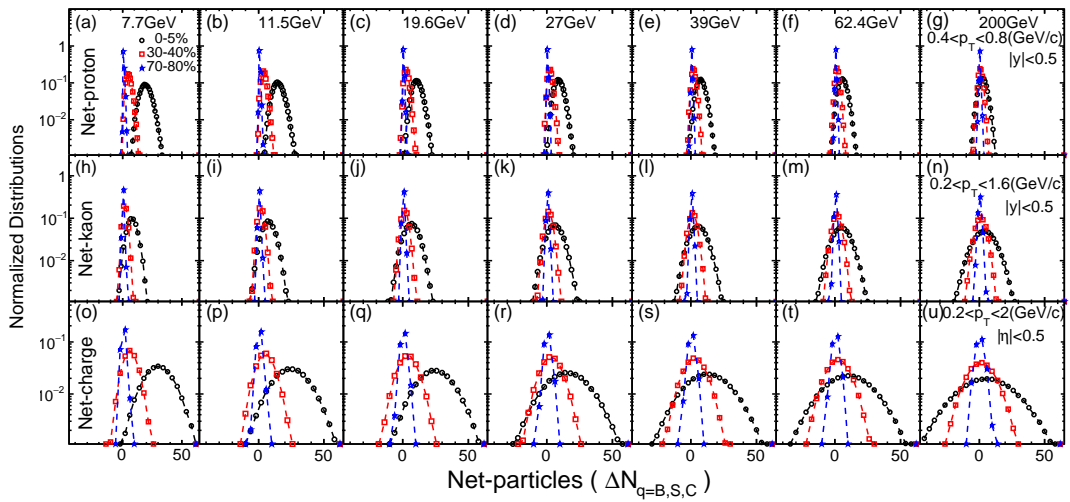}
\caption{\label{Fig_2} (Color online)Event-by-event distributions of net-proton, net-charge and net-kaon multiplicity distributions for Au+Au collisions at $\sqrt{s_{NN}}$ =7.7, 11.5, 19.6, 27, 39, 62.4 and 200GeV for three centrality bins(0-5\%, 30-40\%, 70-80\%). }
\end{figure*}

\begin{figure*}
\hspace{-2.0cm}	
\includegraphics[width=6.in]{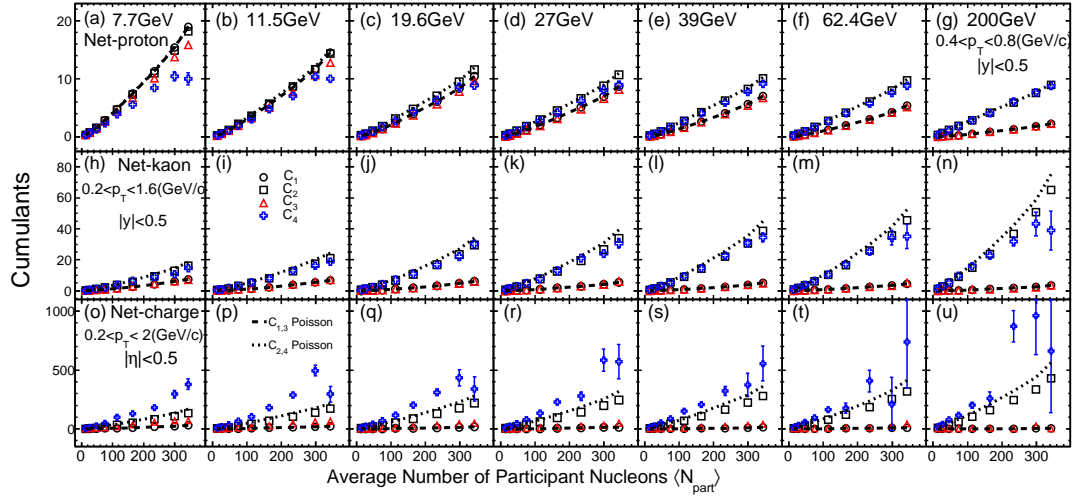}
\caption{\label{Fig_3} (Color online) Centrality dependenece of cumulants ($C_{1} \sim C_{4}$) of net-proton, net-kaon and net-charge multiplicity distributions for Au+Au collisions at $\sqrt{s_{NN}}$ =7.7, 11.5, 19.6, 27, 39, 62.4 and 200GeV. The dashed lines represent the  Poisson expectation. }
\end{figure*}

\parskip=0pt
\begin{figure}[htp]
\includegraphics[width=3.in]{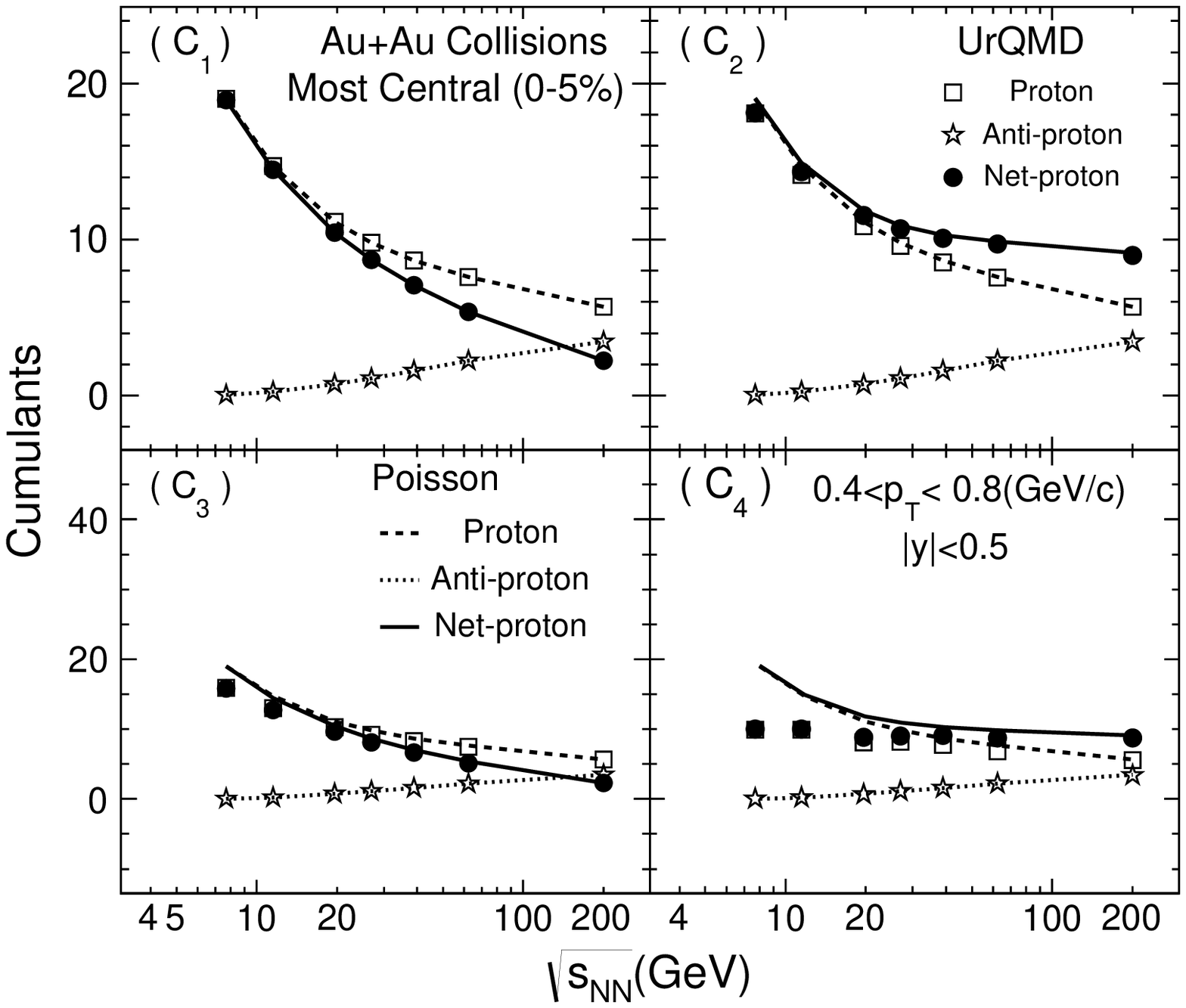}
\caption{\label{Fig_4} Energy dependence of cumulants of proton, anti-proton and net-proton multiplicity distribution for Au+Au collisions at $\sqrt{s_{NN}}$ =7.7, 11.5, 19.6, 27, 39, 62.4 and 200GeV for most central(0-5\%). The lines represent the Poisson expectations.}
\end{figure}

\parskip=0pt
\begin{figure}[htp]
\includegraphics[width=3. in]{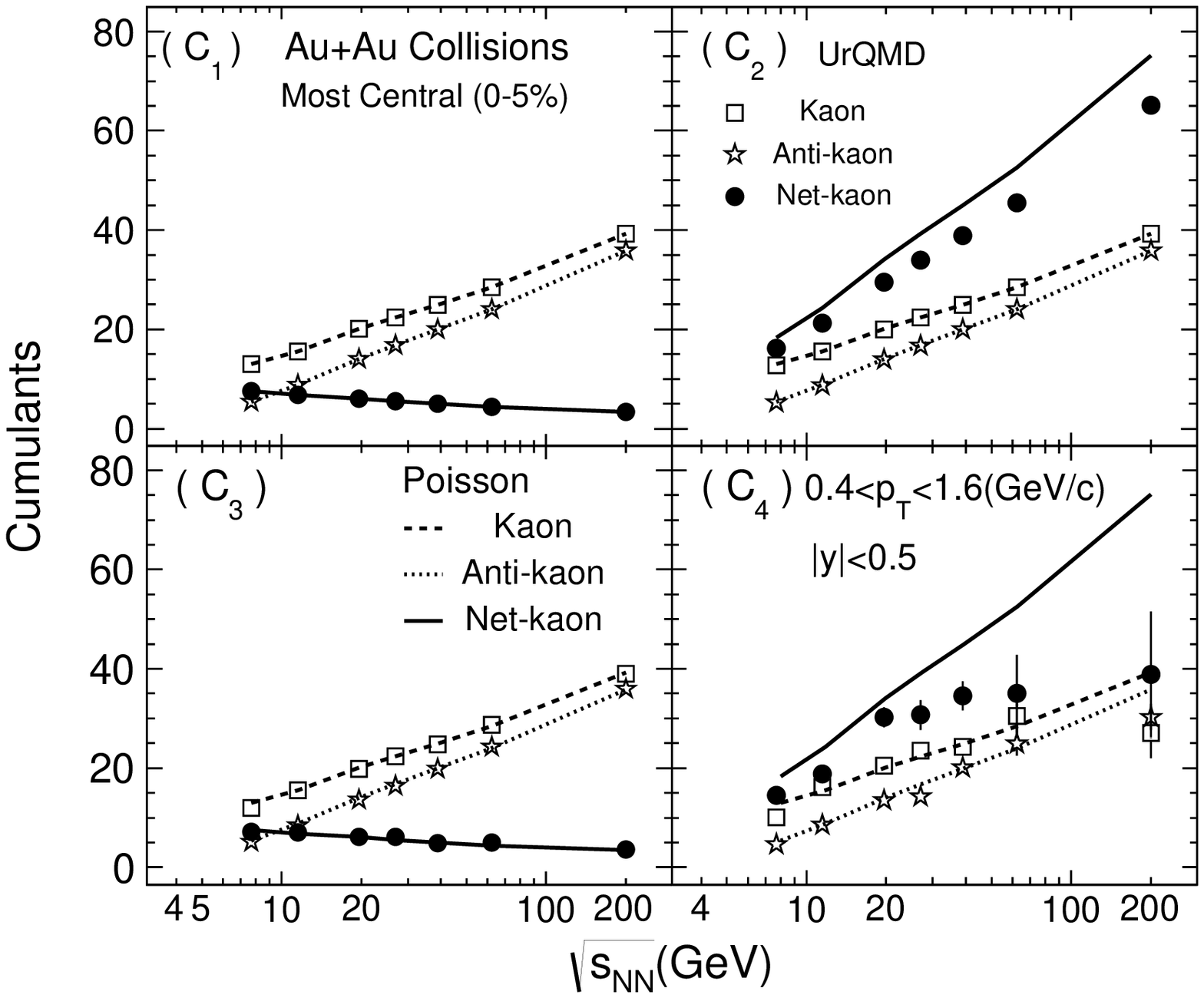}
\caption{\label{Fig_5} Energy dependence of cumulants of kaon, anti-kaon and net-kaon multiplicity distribution for Au+Au collisions at $\sqrt{s_{NN}}$ =7.7, 11.5, 19.6, 27, 39, 62.4 and 200GeV for most central(0-5\%). The lines represent the Poisson expectations.}
\end{figure}

\parskip=0pt
\begin{figure}[htp]
\includegraphics[width=3. in]{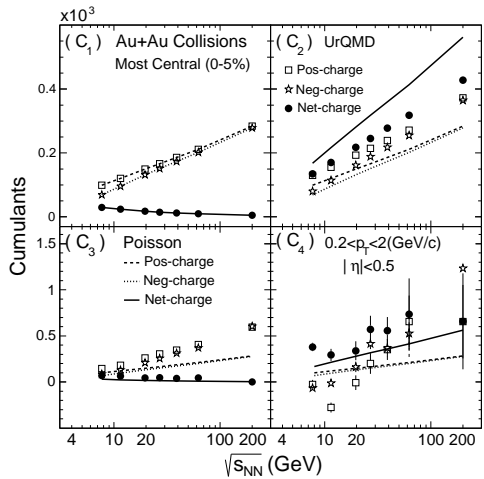}
\caption{\label{Fig_6} Energy dependence of various cumulants of positive charge, negative charge and net-charge multiplicity distribution for Au+Au collisions at $\sqrt{s_{NN}}$ =7.7, 11.5, 19.6, 27, 39, 62.4 and 200GeV for most central(0-5\%). The lines represent the Poisson expectations.}
\end{figure}	

\begin{figure*}
\includegraphics[width=3.4 in]{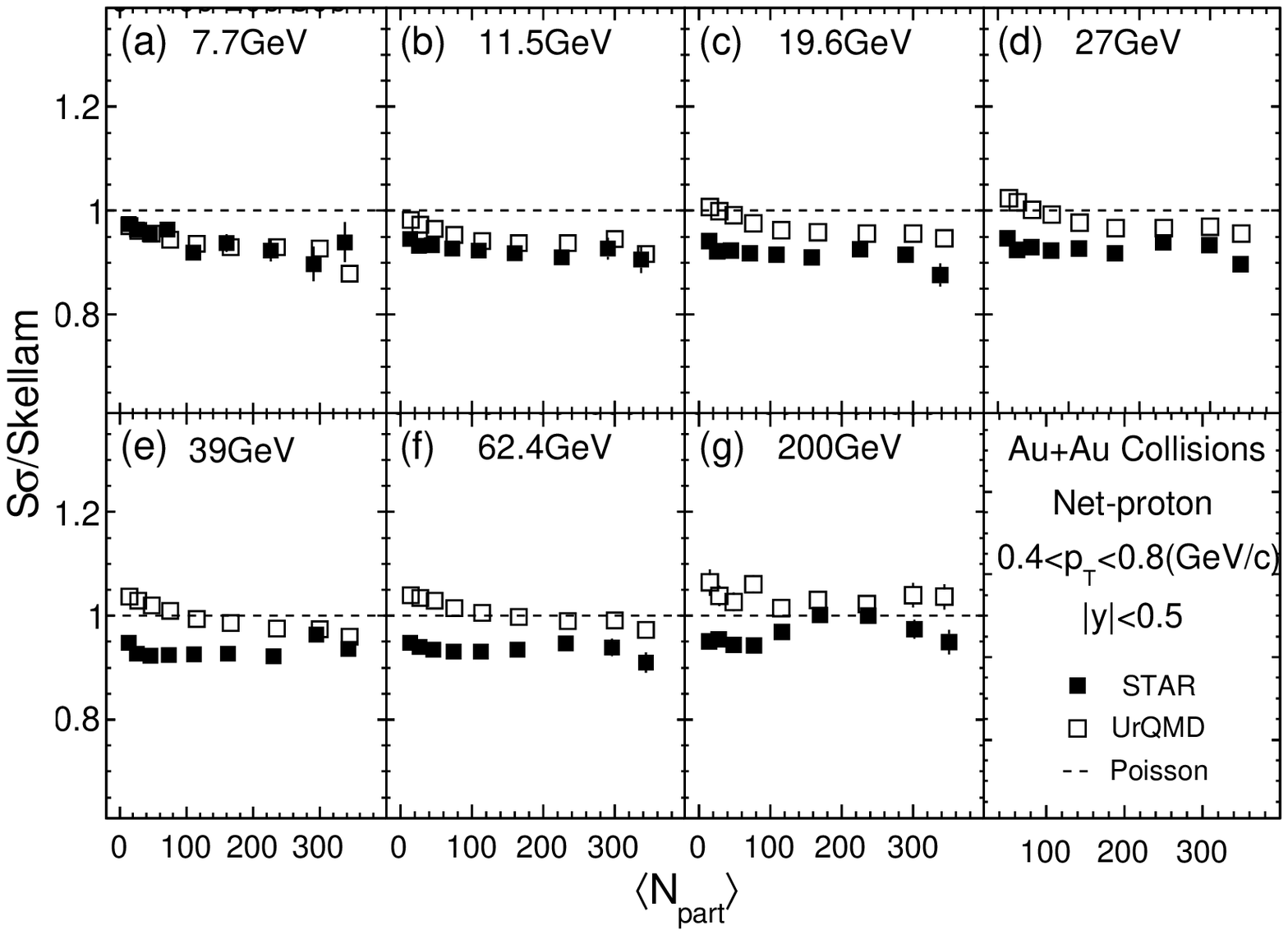}
\includegraphics[width=3.4in]{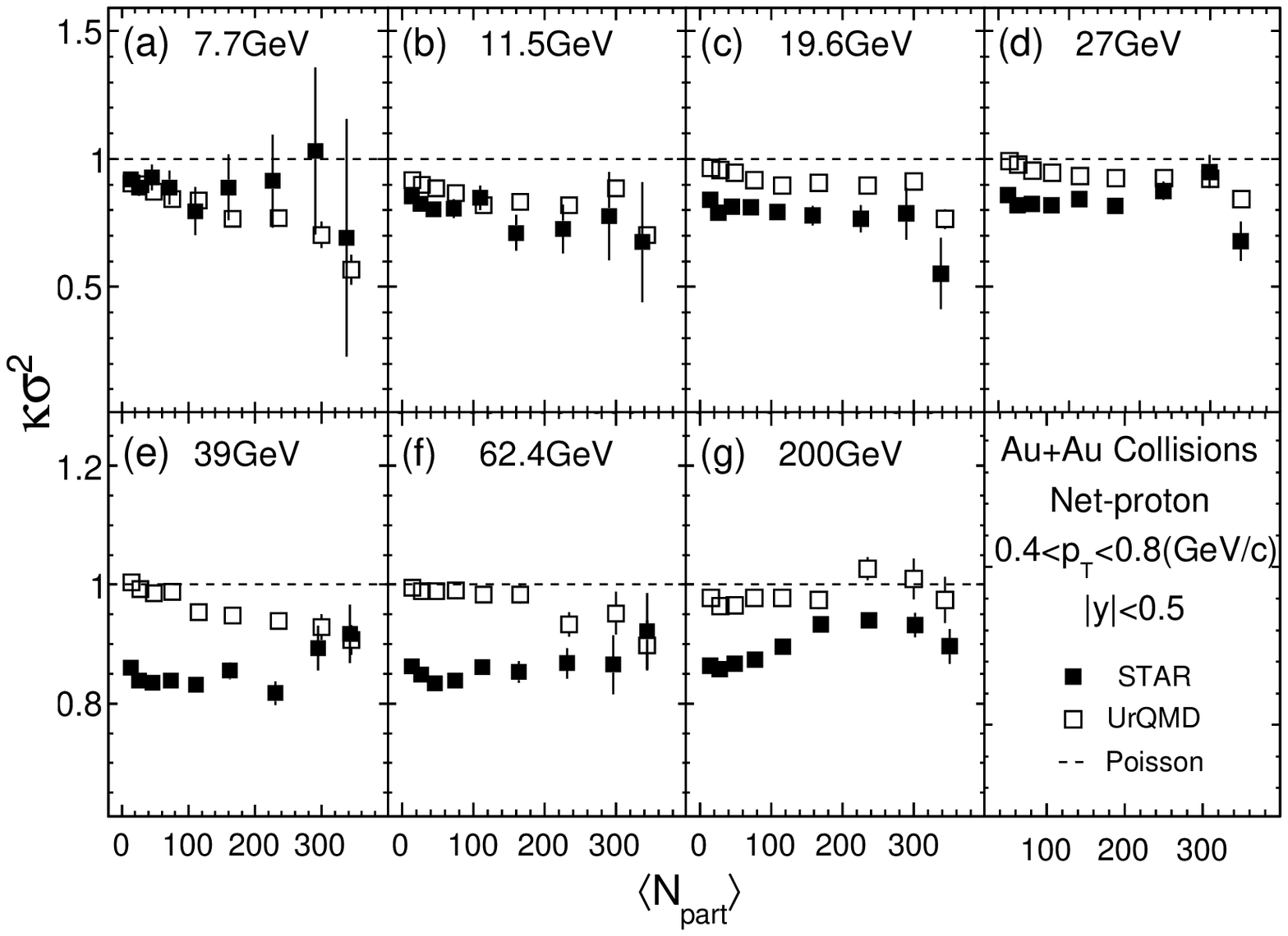}
\caption{Centrality dependence of the cumulant ratios($S\sigma$ , $\kappa\sigma^2$) of net-proton multiplicity distributions for Au+Au collision at $\sqrt{s_{NN}}$ =7.7 to 200 GeV. The solid markers represent the results from STAR measurement, the open markers represent  results from UrQMD calculation.The dashed lines denote the Poisson expectations for the STAR data.} \label{fig:netp_cent}
\end{figure*}

\begin{figure*}
\includegraphics[width=3.4in]{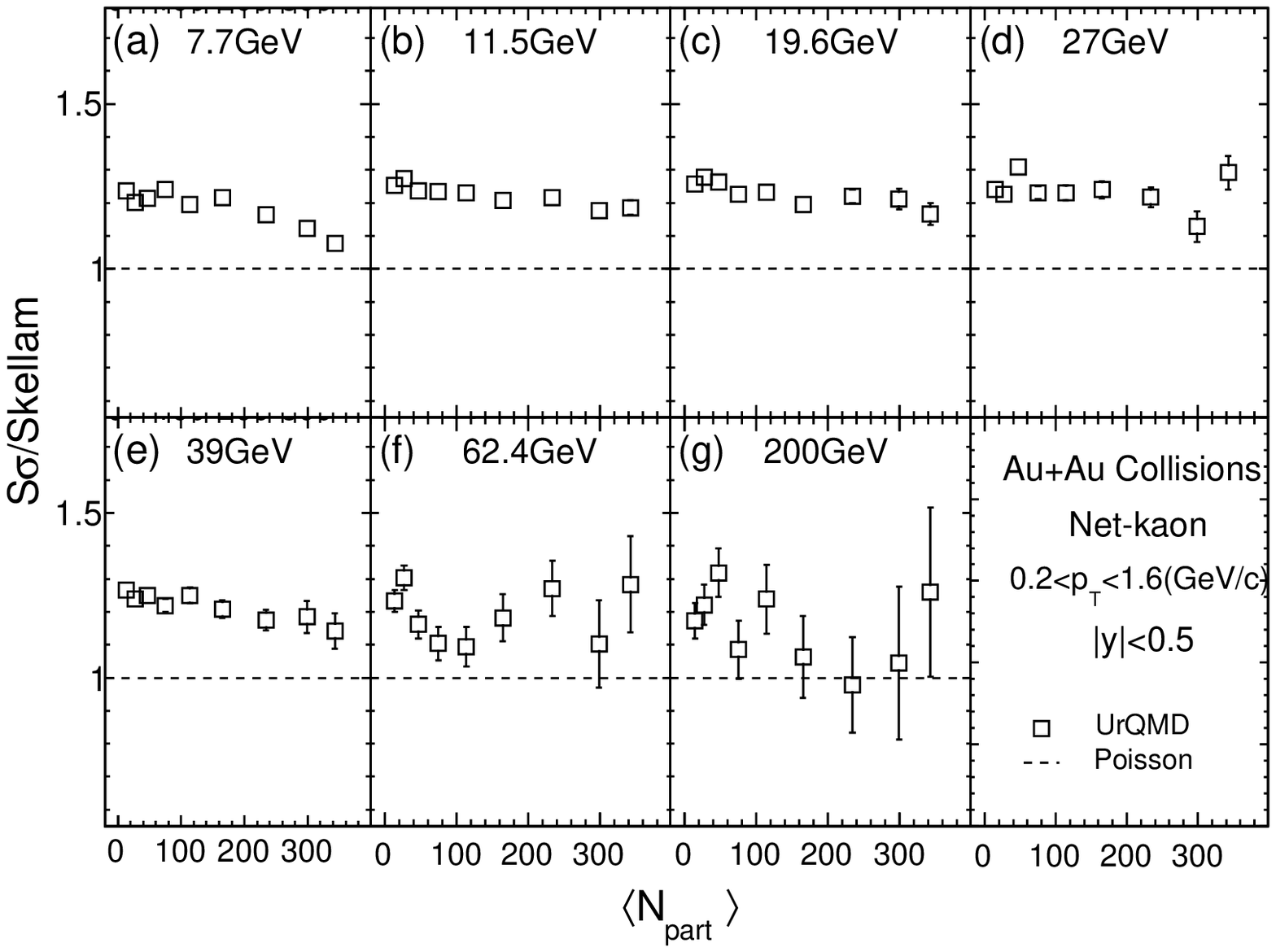}
\includegraphics[width=3.4in]{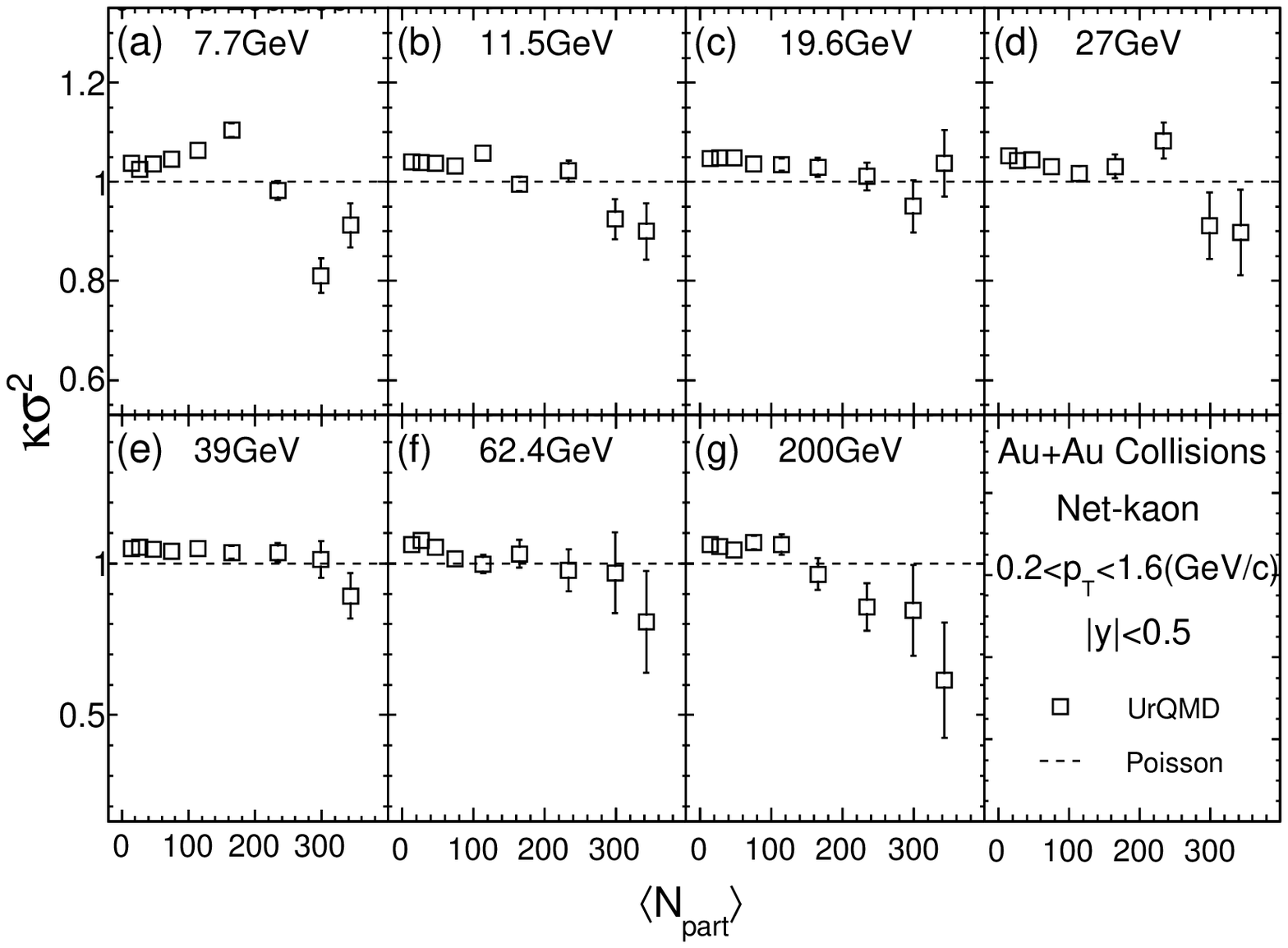}
\caption{ Centrality dependence of the cumulant ratios($S\sigma$ , $\kappa\sigma^2$) of net-kaon multiplicity distributions for Au+Au collision at $\sqrt{s_{NN}}$ =7.7 to 200 GeV. The solid markers represent the results from STAR measurement, the open markers represent  results from UrQMD calculation.}\label{fig:netk_cent}
\end{figure*}

Figure \ref{Fig_1} shows the rapidity distributions ($dN/dy$) of net-proton, proton, anti-proton and net-kaon, $K^{+}$, $K^{-}$,  pseudo-rapidity ($dN/d\eta$) distribution for net-charge, positive charge and negative charge particle multiplicities for the most central  (0-5\%) Au+Au collisions at $\sqrt{s_{NN}}$ = 7.7 to 200 GeV from UrQMD calculations. It is observed that the $dN/dy$ distributions 
of net-proton and proton are nearly flat at mid-rapidity ($|y|<0.5$). The  $dN/dy$ distribution of net-proton closely follow the distributions of proton at low energies and the values at $y=0$ (${\left. {dN/dy} \right|_{y = 0}}$) monotonically increase with decreasing energies. The anti-proton ${\left. {dN/dy} \right|_{y = 0}}$ show monotonically decrease with decreasing collision energies. These can be explained by the interplay between the baryon stopping and pair production of proton and anti-proton from high to low energies. The baryon stopping is stronger at low energies, while the pair production dominate the production of proton and anti-proton at high energies. The $K^+$ and $K^-$ are mainly produced from pair production at high energies, where the number of  $K^+$ and $K^-$ are very similar. At low energies, the associate production of the $K^+$ with a hyperon become more important. Due to electric charge conservation in particle production, the net-charge number will be conserved. Thus, the number of positive charge and negative charge particle multiplicity always follow closely with each other at all energies. 

Figure \ref{Fig_2} shows the event-by-event net-proton, net-charge and net-kaon multiplicity distributions for Au+Au collisions at $\sqrt{s_{NN}}$ =7.7$\sim$200 GeV for three centralities (0-5\%, 30-40\%, 70-80\%). For each energy, the mean value and the width ($\sigma$) of the net-proton, net-charge and net-kaon distributions are larger for central than peripheral collisions.  For the same centrality, the mean value of net-proton, net-kaon and net-charge multiplicity distributions become larger as decreasing the collision energy. The width of the net-kaon and net-charge multiplicity distributions decrease with decreasing the collision energy while the net-proton distribution shows opposite trend.  The width of the net-charge distributions are much larger than the net-proton and net-kaon distribution.  Based on the Delta theorem \cite{ref18,ref18_1}, the statistical errors of various cumulants are proportional to the different power of $\sigma$ value of the distributions. Thus, the cumulants of net-charge multiplicity distributions are with larger statistical errors than the net-proton and net-kaon cumulants with the same number of events. On the other hand, the volume fluctuations originated from the 
finite centrality bin width and initial volume (geometry) fluctuations of the colliding nuclei \cite{ref19} will enhance the fluctuation measurements. The raw multiplicity distributions of net-proton, net-charge and net-kaon shown in Fig.\ref{Fig_2} should not be directly used to calculate the various cumulants and one needs to apply so called centrality bin width correction \cite{ref18} to address the effects of the volume fluctuations in a wide centrality bin. 

Figure~\ref{Fig_3} shows the centrality dependence of cumulants (up to fourth order) of net-proton, net-charge and net-kaon multiplicity distributions in Au+Au collisions at $\sqrt{s_{NN}}$ =7.7$\sim$200 GeV. In general, those cumulants show a linear variation with the averaged number of participant nucleons, which can be understood as the additivity property of the cumulants by increasing the volume of the system. The odd order cumulants ($C_1$ and $C_3$) and the even order cumulants ($C_2$ and $C_4$) are separated into two groups as the $C_{1}$ and $C_{2}$ values are closely follow the $C_{3}$ and $C_{4}$, respectively. If the multiplicity distributions of particles and anti-particles are independent Poissonian, the corresponding net-particle cumulants can be simply constructed from the mean values as 
\begin{equation}  \label{eq:pos} 
{C_n}(X-Y)= {C_1}(X) + {( - 1)^n}{C_1}(Y)
\end{equation} 
where $C_{1}(X)$ denotes the mean value of proton, $K^{+}$ and positive charge particle distributions, $C_{1}(Y)$ represents the mean value of anti-proton, $K^{-}$ and negative charge particle distributions, respectively.  The separation gets smaller when the energy decreases because of the reduction of number of anti-proton and $K^{-}$ produced from pair production at low energies. 

\begin{figure*}
\includegraphics[width=3.6in]{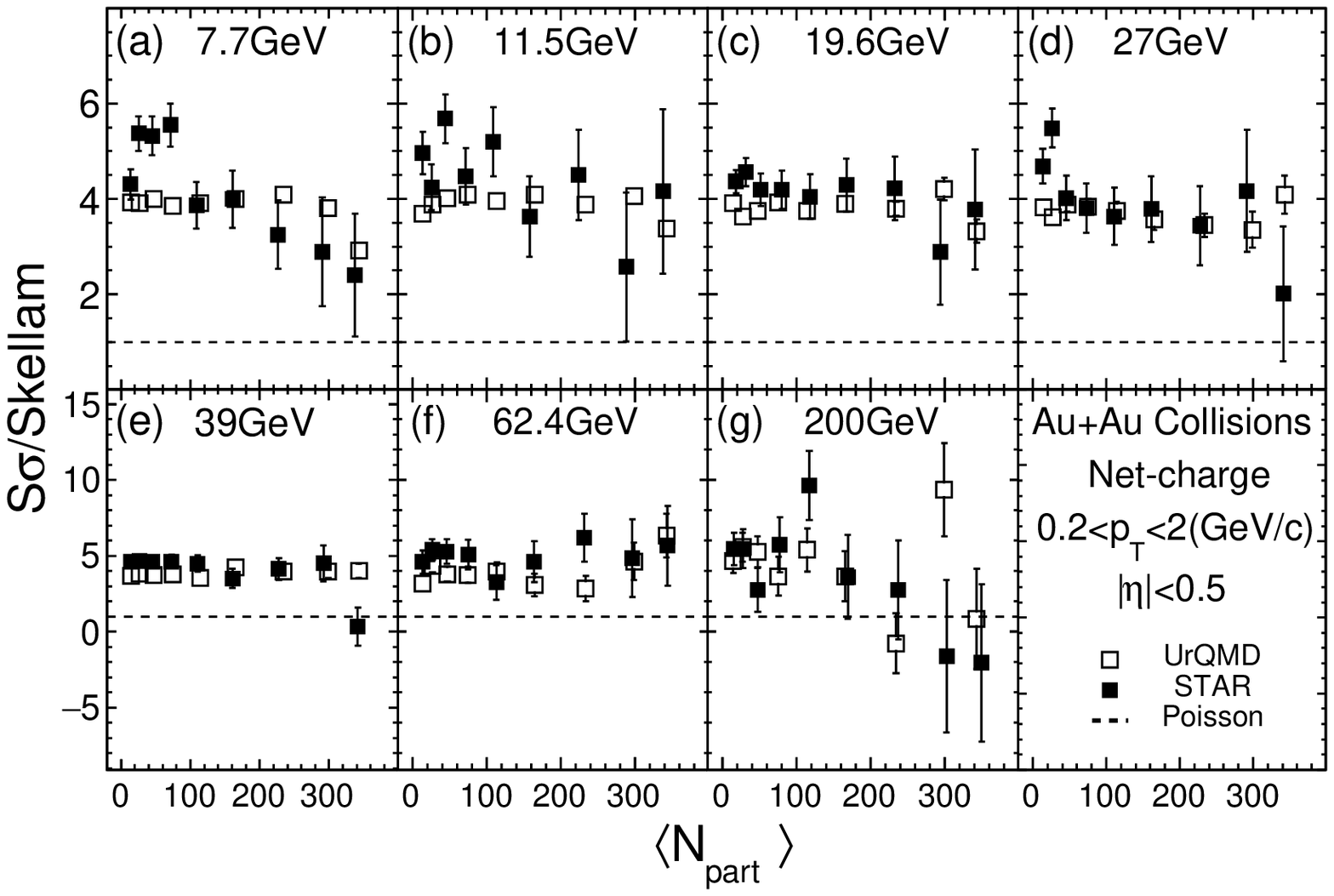}
\includegraphics[width=3.3in]{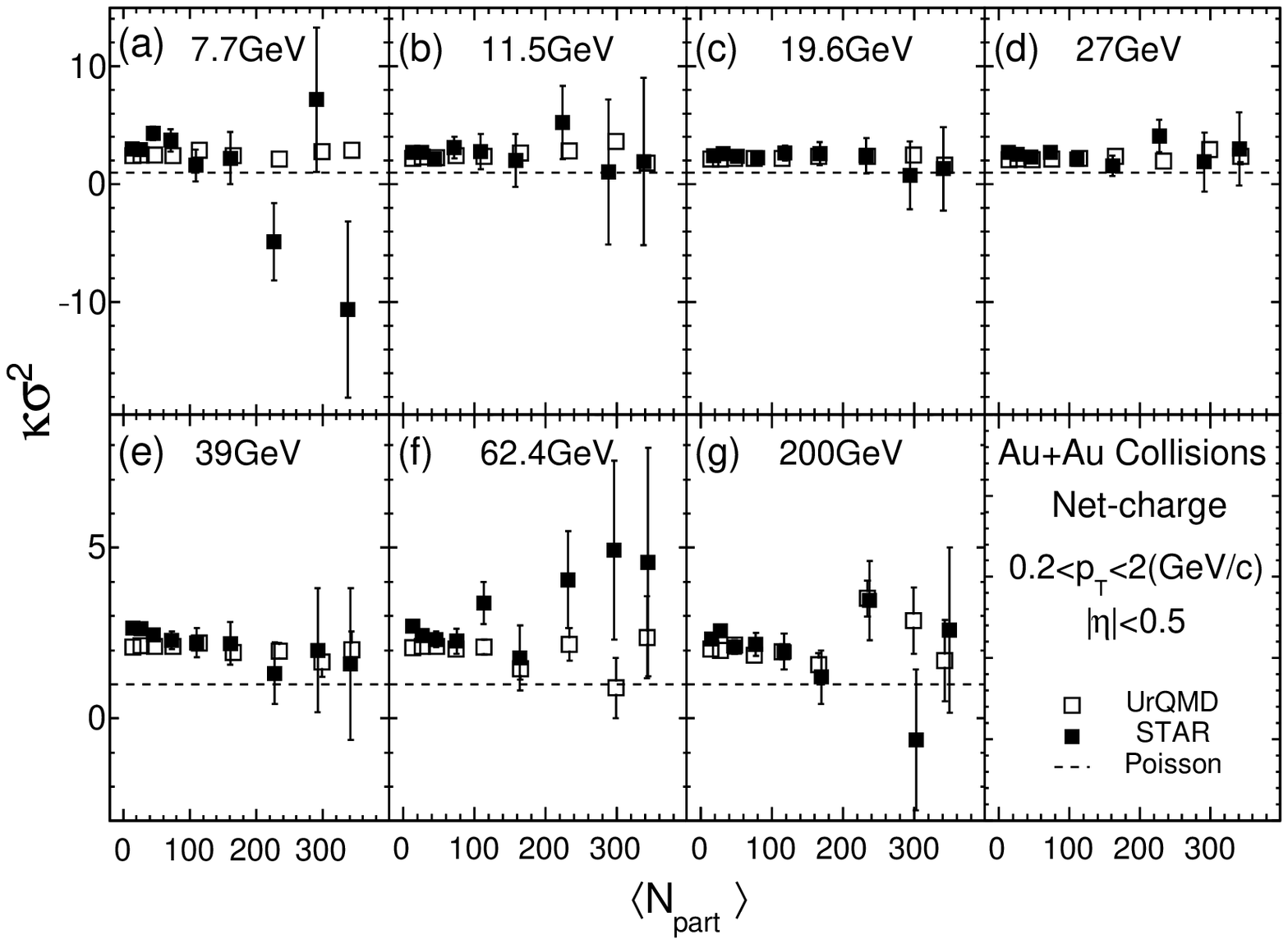}
\caption{ Centrality dependence of the cumulant ratios($S\sigma$ , $\kappa\sigma^2$) of net-charge multiplicity distributions for Au+Au collision at $\sqrt{s_{NN}}$ =7.7 to 200 GeV. The solid markers represent the results from STAR measurement, the open markers stand for the results from UrQMD calculation. The dashed lines denote the Poisson expectations for the STAR data.} \label{fig:netq_cent}
\end{figure*}
Figure \ref{Fig_4} shows energy dependence of cumulants ($C_{1} \sim C_{4}$) of proton, anti-proton and net-proton multiplicity distributions for the most central(0-5\%)  Au+Au collisions at $\sqrt{s_{NN}}$ =7.7 to 200 GeV.  The cumulants of net-proton and proton distributions monotonically increase when energy decreases, while the cumulants of anti-proton distribution show decreasing trends. The net-proton cumulants follow closely with the proton cumulants at low energies. Those can be understood by the interplay between baryon stopping and pair production at different energies. The  anti-proton cumulants ($C_{1} \sim C_{4}$)  can be well described by the Poisson expectations, while large deviations from Poisson expectations are observed for the $C_{3}$ and $C_{4}$ of net-proton and proton multiplicity distributions, especially for the low energies. Figure~\ref{Fig_5} shows the energy dependence of cumulants of the net-kaon, $K^{+}$ and $K^{-}$ distributions. The mean value of $K^{+}$ distributions decrease with decreasing energies, which is opposite to the trend of the mean value of protons shown in Fig.\ref{Fig_4}.  At low energies, baryon stopping dominated the production of protons at mid-rapidity. However, the $K^{+}$ and $K^{-}$ are produced particles and the pair production is the main production mechanism, which leads to the monotonically increase of the yields when the collision energy increases. Furthermore, the odd order cumulants ($C_{1}$ and $C_{3}$) and the even order cumulants ($C_{2}$ and $C_{4}$) of net-kaon distributions shows opposite energy dependence trend.  Those can be explained qualitatively in terms of the Poisson expectations of the cumulants in Eq. \ref{eq:pos} and the energy dependence of the associate and pair production mechanism for the $K^{+}$ and $K^{-}$.  When the energy increases, the pair production become more important and the yields of $K^{+}$ and $K^{-}$ increase with increasing energy. That's why the even order cumulants ($C_{2}$ and $C_{4}$) of net-kaon distributions show monotonically increase trends. On the other hand, the associate production of $K^{+}$ will increase when the energy decrease. The $C_{2}$ of net-kaon multiplicity distributions deviates below the Poisson baselines, while the $C_{3}$ of net-kaon distributions are better fitted with the Poisson values. To understand this, we derived the formula of $C_{2}$ and $C_{3}$ of net-particles in term of the cumulant of particle and anti-particles, respectively. Thus, we have: 
\begin{equation}
{C_2}(X - Y) = {C_2}(X) + {C_2}(Y) - 2 \times Cov(X,Y), \label{eq:C2_corr}
\end{equation}
\begin{equation}
\begin{split}
{C_3}(X - Y) &= {C_3}(X) - {C_3}(Y)\\
 &+ 6 \times [ < X >  -  < Y > ] \times Cov(X,Y)\\
 &- 3 \times [Cov({X^2},Y) - Cov(X,{Y^2})],   \label{eq:C3_corr}
\end{split}
\end{equation}
where the $X$ and $Y$ represent the particle and anti-particle multiplicities, respectively. For e.g, number of protons, anti-protons and $K^{+}$ and $K^{-}$. Since the  $C_{2}$ and $C_{3}$ of the  $K^{+}$ and $K^{-}$ multiplicity distributions can be well described by the Poisson baselines, respectively ($C_{2}(X)=C_{1}(X )$, $C_{2}(Y)=C_{1}(Y )$ ), the deviation of the net-kaon $C_{2}$ below Poisson values indicates a large positive covariance between $K^{+}$ and $K^{-}$. In other words, there is a strong positive correlation between $K^{+}$ and $K^{-}$. However, it is not quite straightforward to understand why it shows better description of $C_{3}$ of net-kaon by Poisson expectations. In the Eq.~\ref{eq:C3_corr}, the correlation term is different from the one in the Eq.~\ref{eq:C2_corr} and in addition to the covariance between  $K^{+}$ and $K^{-}$,  there is another term, $ Cov({X^2},Y) - Cov(X,{Y^2})$. Apparently, if the term $6(< X> - < Y> )Cov(X,Y)- 3 (Cov(X^2,Y) - Cov(X,Y^2))$ is close to zero for $X= K^{+}$  and $Y= K^{-}$ multiplicity ,  we obtain the ${C_3}(X - Y)\approx{C_3}(X) - {C_3}(Y)$, which certainly can be well described by Poisson baselines. The cumulants of net-charge, positive and negative charge multiplicity distributions are shown in  Fig. \ref{Fig_6}.  The energy dependence trends for cumulants in Fig. \ref{Fig_6} are similar in Fig. \ref{Fig_5}, since most of the charged particles (e.g. kaon and pion) are directly produced. The deviations from the Poisson expectations begin to show up at second order cumulants ($C_{2}$). It indicates a strong correlation between the positive and negative charged particles in heavy-ion collisions. It is for the similar reason as for the net-kaon shown in the Fig.~\ref{Fig_5} that the $C_3$ of net-charges fit better with the Poisson baselines than $C_{2}$. 

In the above discussion, we may find that the different production mechanism of different particles species have a significant impact on the cumulant of net-particles distributions and with a strong energy dependence. For. e.g., the proton and anti-proton are mainly produced by pair-production at high energies, while at low energies the baryon stopping of incoming nucleon becomes dominating production source for the protons and anti-protons. On the other hand, the charged kaon and pion are mainly directly produced in the heavy-ion collisions, and their yields will monotonically decrease as the collision energy decreases. Due to the $K^{+}$ associate production with hyperon,  the yield of $K^{+}$ is higher than that of $K^{-}$ especially at low energies where the baryon density is larger.

%%%%%%%%%%%%%%%%% (5) Results comparison with STAR experiment %%%%%%%%%%%%%%%%%%%%%

\section{Comparisons Between STAR Data and UrQMD Model}
The STAR Collaboration has published the centrality and energy dependence of the moments of net-proton and net-charge multiplicity distributions for Au+Au collision at $\sqrt{s_{NN}}$=7.7, 11.5, 19.6, 27, 39, 62.4, 200 GeV~\cite{ref17.1,ref20}. Those data are taken from the first phase of RHIC BES program in the years 2010 and 2011 \cite{ref23}. Recently, the new net-proton and net-kaon results have been presented in CPOD2014 and QM2015 conference~\cite{CPOD2014_XFLUO,QM2015_XFLUO,QM15_netkaon}, respectively. The net-proton fluctuation measurement $\kappa\sigma^2$ of most central Au+Au collisions clearly show non-monotonic energy dependence after extending the transverse momentum ($p_{T}$) coverage of proton and anti-proton from $0.4\sim0.8$ to $0.4\sim2$ GeV/c. The acceptance dependence of the fluctuation measurements have been discussed by the theoretical calculations in terms of critical contribution and the thermal blurring/diffusion effects in heavy-ion collisions~\cite{acc1,acc2}. In the following, we will make a comparison for the cumulant ratios (\SD\ and \KV ) of the net-proton, net-charge and net-kaon distributions between the STAR data and UrQMD calculations.

\begin{figure*}
\includegraphics[width=3.in]{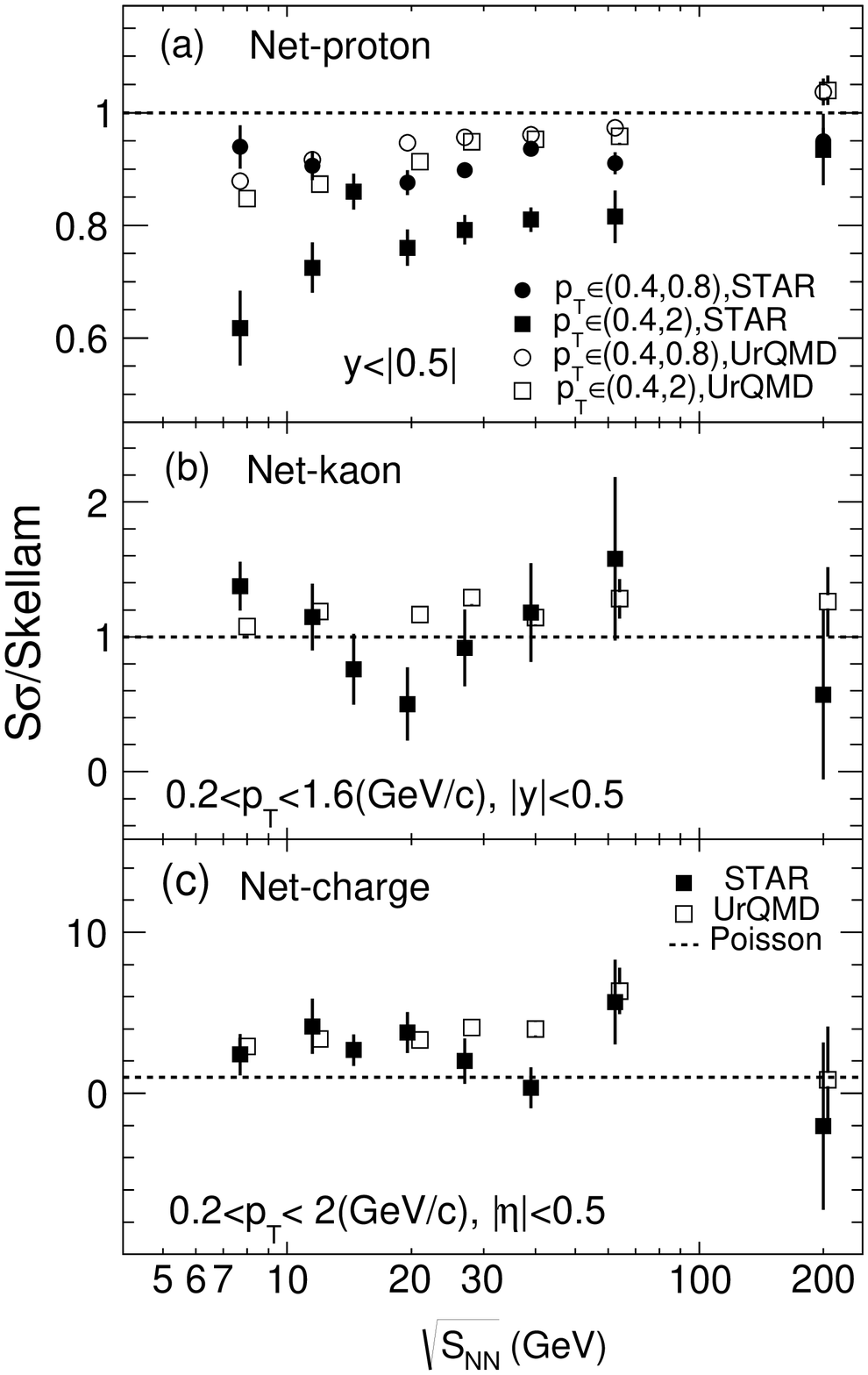}
\includegraphics[width=3.in]{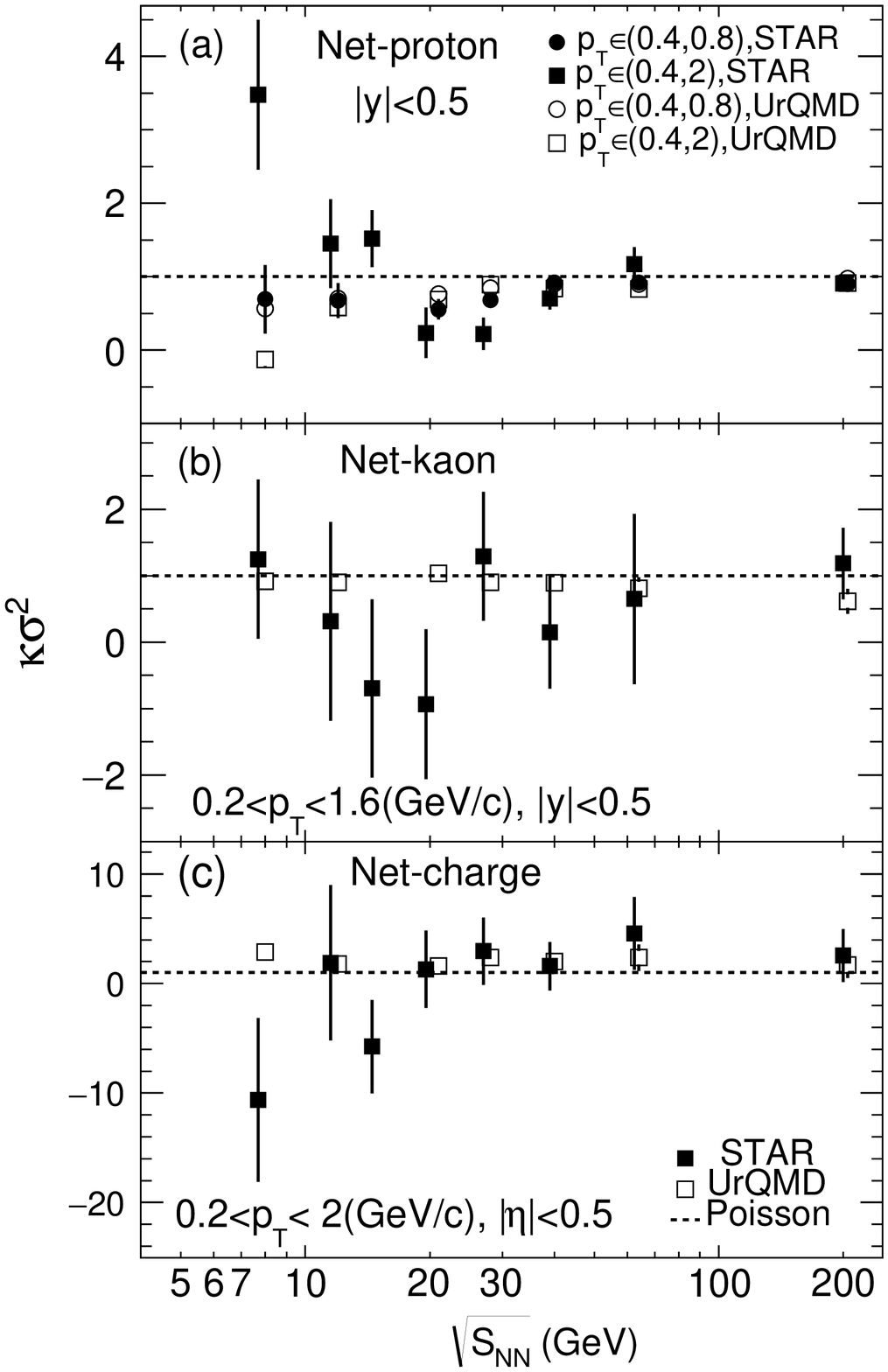}
\caption{Energy dependence of cumulant ratios($S\sigma$, $\kappa\sigma^2$) of net-proton, net-charge and net-kaon multipliity distributions for Au+Au collision at $\sqrt{s_{NN}}$ =7.7 to 200 GeV. The solid markers represent the results from STAR measurement, the open markers represent  results from UrQMD calculation. The dashed lines denote the Poisson expectations for the STAR data.} \label{fig:erg_SD_KV} 
\end{figure*}

Experimentally, the volume independent cumulant ratios ($C_{3}/C_{2}=S\sigma$, $C_{4}/C_{2}=\kappa\sigma^2$) are constructed to be the main observables to search for the QCD critical point. Fig. \ref{fig:netp_cent} shows the centrality dependence of $S\sigma$/Skellam and $\kappa\sigma^2$ of net-proton multiplicity distributions for Au+Au collisions from $\sqrt{s_{NN}}$=7.7 to 200GeV.  Based on the Eq. \ref{eq:pos}, the Skellam (Poisson) baseline for net-proton \SD\ is constructed from the mean value of the proton ($<N_{p}>$) and anti-proton ($<N_{\bar{p}}>$) number as $(<N_{p}>-<N_{\bar{p}}>)/(<N_{p}>+<N_{\bar{p}}>)$. For $S\sigma$/Skellam and {\KV},  the UrQMD results follow closely with the STAR measurements at 7.7 and 11.5 GeV whereas  the experimental data show larger deviation from the Poisson baselines than the model results.  In fig. \ref{fig:netk_cent}, as the centrality dependence of net-kaon cumulant ratios from BES energies is not public yet,  we only show the results from the UrQMD model calculations. It is observed that the $S\sigma$/Skellam values are always above unity, which means $S\sigma$ values are larger than Poisson baselines. The net-kaon {\KV} from UrQMD calculations are close to unity. Figure \ref{fig:netq_cent} shows the centrality dependence of cumulants ratios  ($S\sigma$, $\kappa\sigma^2$) of net-charge multiplicity distributions. The $S\sigma$ from UrQMD calculations are in general smaller than the STAR data, which show large deviation from the Poisson expectation at low energies. For net-charge $\kappa\sigma^2$, the STAR results can be well described by the UrQMD results, however the experimental results are with large statistical errors.  

Figure \ref{fig:erg_SD_KV} shows the energy dependence of cumulant ratios ($S\sigma, \kappa\sigma^2$) of net-proton, net-charge and net-kaon multiplicity distributions of the 0-5\% most central Au+Au collisions at RHIC BES energies from the STAR experiments and UrQMD calculations. The \SD\ values have been scaled by the Skellam (Poisson) baselines. Within statistical uncertainties, the net-charge and net-kaon results has weak energy dependence and are close to Skellam (Poisson) baselines. Both of the net-charge and net-kaon $S\sigma, \kappa\sigma^2$ measured by STAR experiment can be well described by the UrQMD model. On the other hand, the STAR experiment has measured the high order net-proton fluctuations for the two transverse momentum ({\pT}) coverages $0.4\sim0.8$ GeV/c and  $0.4\sim2$ GeV/c, respectively. For the net-proton $S\sigma$/Skellam, both the UrQMD results and STAR data show monotonic decreasing trends as the collision energy decreases. For net-proton $\kappa\sigma^2$ with the small  \pT\ coverage $0.4\sim0.8$ GeV/c,  the STAR data show large deviations from unity below \sNN\ = 39 GeV, especially at 19.6 and 27 GeV. The statistical errors of the data at 7.7 and 11.5 GeV are much larger than higher energies due to the limited statistics.  For wider \pT\ coverage $0.4\sim2$ GeV/c, we observe a clear non-monotonic energy dependence for the net-proton $\kappa\sigma^2$ with a minimum around 19.6 GeV, then
become above unity in the energies below 19.6 GeV.  However, the UrQMD model without implementing the critical point physics,  fails to describe the STAR data especially for large \pT\ coverage $0.4\sim2$ GeV/c and low collision energies. The monotonic decrease when decreasing energies and strong suppression below unity at low energies are consistent with the effects of the baryon number conservations.

%%%%%%%%%%%%%%%%%%% (6) Summary %%%%%%%%%%%%%%%%%%%%%%

\section{Summary}
Experimentally, fluctuations of conserved quantities have been applied to probe the signature of the QCD phase transition and critical point in heavy-ion collisions.
To understand the non-critical contributions to the observables, we have performed detailed model calculations. In this paper, we present the centrality and energy dependence of the cumulants ($C_{1}\sim C_{4}$) and their ratios ($C_{3}/C_{2}=S\sigma, C_{4}/C_{2}=\kappa\sigma^2$) of net-proton, net-charge and net-kaon multiplicity distributions with UrQMD model for Au+Au collision at $\sqrt{s_{NN}}$=7.7, 11.5, 19.6, 27, 39, 62.4 and 200 GeV. The production mechanisms of the proton and kaon have a significant impact on the fluctuations of net-particles. For e.g, the interplay of the baryon stopping, pair production of proton and anti-proton, and the associate, pair production of the $K^{+}$ and  $K^{-}$ at different energies.  At low energies, the baryon stopping of protons and associate production of kaons play an important roles. Those will lead to big differences between the cumulants of particles and anti-particles distributions, such as proton, anti-protons and $K^{+}$,  $K^{-}$.  Finally, the comparisons for the cumulant ratios ($S\sigma$, $\kappa\sigma^2$) of net-proton, net-charge and net-kaon multiplicity distributions have been made between the STAR data and the UrQMD calculations. Within the statistical uncertainties, the net-charge and net-kaon fluctuations measured by STAR experiment can be described by the UrQMD results. For the net-proton fluctuations, the STAR measured \KV\ at 0-5\% most central Au+Au collisions show a clear non-monotonic energy dependence with a minimum around 20 GeV. This non-monotonic behavior can not be described by the UrQMD model, in which there has no critical physics implemented. The large suppression of the net-proton fluctuations at low energies could be explained by the effects of baryon number conservations. Although the physics of the QCD critical point is not implemented in the UrQMD simulation, the results from UrQMD calculations can provide us non-CP physics baselines and a qualitative estimation for the background contributions to the QCD critical point search in heavy-ion collisions by using the fluctuations of the net-proton, net-kaon and net-charge numbers.

%%%%%%%%%%%%%%%%%%% Acknowledgments %%%%%%%%%%%%%%%%%%
\section*{Acknowledgments}
The authors thank Dr. Jochen Thaeder and Prof. Nu Xu for the constructive and helpful discussion. 
J. Xu was supported by the China Scholarship Council (CSC) and Lawrence Berkeley National Lab (LBNL). The authors from CCNU also thank the supports from the STAR Collaboration. The work was supported in part by the MoST of China 973-Project No.2015CB856901, NSFC under grant No. 11575069, 11221504.

%%%%%%%%%%%%%%%%%% Citation %%%%%%%%%%%%%%%%%%%%%%%%%

\ed